\input harvmac
\input epsf
\lref\angelica{A.~de~Oliveira-Costa and G.F.~Smoot, Ap.\ J.\ {\bf 448}, 447 (1995).}
\lref\wittenknot{E.~Witten, Commun.\ Math.\ Phys.\  {\bf 121}, 351 (1989).}
\lref\tschierske{C.~Tschierske, Curr.\ Opin.\ Colloid.\ In.\ {\bf 7}, 69 (2002).}
\lref\ppm{W.~Cao, A.~Mu\~noz, P.~Palffy-Muhoray and B.~Taheri, Nature Materials {\bf 1}, 111 (2002).}
\lref\weitz{The Article}
\lref\warner{H.~Finkelman, E.~Nishikawa, G.G.~Pereira and M.~Warner, Phys.\ Rev.\ Lett.\ {\bf 87}, 015501 (2001).}
\lref\percec{V.~Percec, M.~Glodde, T.K.~Bera, Y.~Miura, I.~Shiyanovskaya, K.D.~Singer, V.S.K.~Balagurusamy, P.A.~Heiney, I.~Schnell, A.~Rapp, H.-W.~Spiess, S.D.~Hudson and H.~Duan, Nature {\bf 419}, 384 (2002).}
\vbox{\baselineskip=0.18truein
\centerline{\bf Topology from the Bottom Up}
\vskip10pt
\centerline{Randall D. Kamien}
\centerline{\sl Department of Physics and Astronomy, University of Pennsylvania,}
\centerline{\sl Philadelphia, PA 19104-6396 USA}}

Though the use of topology and geometry to understand the physical world is commonplace and natural it is only now that it has become central to the control and design of new materials.  For instance, data from the cosmic
microwave background explorer (COBE) has not only shed light on the geometry of the Universe and the cosmological constant, but it constrains the allowed topology as well \angelica, ruling out the possibility of the its being spatially periodic.  The classical theory of electromagnetism can be recast geometrically and serves as the simplest example of the gauge theories of elementary particles.  These  theories are the basis for the wildly successful standard model of particle physics; their deep connection to modern topology has led to major advances in pure mathematics \wittenknot.  

In the realm of condensed matter and materials physics, the role of topology is well known.  For instance, the number of times that a closed string winds around a point is independent of its exact conformation and is therefore a topological property. The persistence of currents in superconductors and superfluids is guaranteed by this very same topology as is the existence of Abrikosov flux lines in Type II superconductors and defects in smectic liquid crystals.  These topological defects are at best a profound example of a macroscopic quantum state and at worst a nuisance for technology.  On page 1716 of this issue Bausch {\sl et al.}\ref\bausch{A.R.~Bausch, M.J.~Bowick, A.~Cacciuto, A.D.~Dinsmore, M.F.~Hsu, D.R.~Nelson, M.G.~Nikolaides, A.~Travesset and D.A.~Weitz, Science {\bf 299}, 1716 (2003).} have performed a series of elegant experiments to convincingly confirm that the robust and generic aspects of the topology of a sphere or soccer ball can be used to understand the structure of colloidal crystals on spheres.  Not just a curiosity, these structured colloidosomes will be the building blocks for complex structures with novel optolectronic, mechanical and photonic properties.

Through a variety of methods, materials physics has developed the tools to predict and control the assembly of complex materials.  The most promising are methods based on self-assembly, those that rely on identical molecules and supramolecular assemblies to assemble into structures that reflect their molecular architecture \tschierske.   Typically, advances in chemical synthesis lead to ever more complex molecules that, in turn, assemble into ever more complex structures.  Perhaps the simplest example of this is a chiral nematic or cholesteric liquid crystal.  In those materials control of the rigidity, aspect ratio and chirality of the molecules allows for the self-assembly of a singly-periodic cholesteric phase, {\sl i.e.} a one-dimensional crystal.   This level of macroscopic structure is enough to make temperature probes, field-controlled diffraction gratings, and optical switches.  The chiral molecular structure is often enough to form exotic liquid-crystalline blue phases that control light in three-dimensions and exhibit mirrorless lasing \ppm.
\vskip10pt
\centerline{\vbox{\hsize=5truein \epsfxsize=5truein\baselineskip=0.18truein
\centerline{\epsfbox{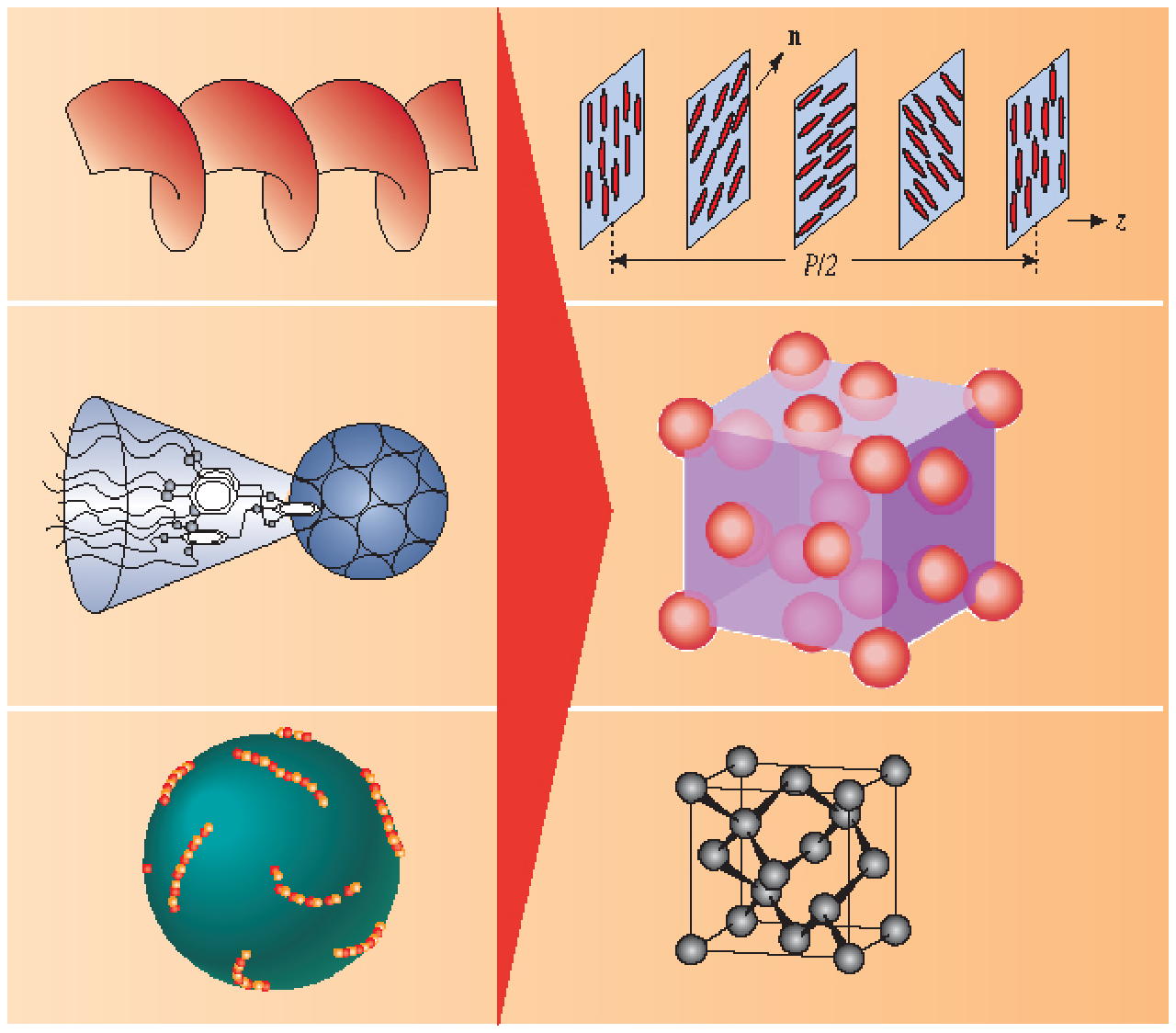}}
\vskip2pt
{\noindent The control of molecular geometry and topology can lead to the rational design and control of
macroscopic structure.  The cholesteric phase of chiral mesogens and the $Pm\overline{3}n$ phases of
dendrimers are the precursors to the new phases which can be built from scarred colloidosomes.}
}}

The ability to control the topology of the molecules in addition to their geometry leads to concomitant advances in the spectrum of materials properties.  Polymers with attached, side-chain nematogens form liquid crystalline elastomers \warner\ with extraordinary opto-mechanical responses, exhibiting shape changes between 10\% and 400\%!  Thus, by merely changing the connectivity, {\sl i.e.} the topology, of the mesogens and the polymer backbones, it is possible to drastically alter material properties. These materials are likely to find their way into a variety of actuators and sensors, including artificial muscle.  Another example of topological control is in the synthesis of branched, dendritic polymers or dendrimers.  By controlling the geometry of the branching points and the recursive topology of the branches, it has become possible to assemble supramolecular, nanoscale structures with specific form and function.  Indeed,  recent advances in dendrimer synthesis have led to an entirely new class of complex, electronic materials that owe their great technological potential to their ability to self-process themselves into the nanoscale geometries that modern molecular electronics requires \percec.  Further advances promise mechanical actuators, chemical sensors and anisotropic conductors.

The work by Bausch {\sl et al.} opens up an entirely new mode for self-assembly.  By relying on the elegant topological arguments that ensure twelve scars on each colloidosome, the authors have demonstrated that micron-sized building blocks can, themselves be self-processed.  
The 
universal behavior that is independent of the colloidal interactions makes it possible to design macroscopic materials without the need for elegant, difficult yet costly synthetic methods.
Moreover, since the length of the scars can be rationally controlled, it is possible to tailor the surface properties to enhance or inhibit the binding of specific agents.  The interactions that control the positions and distribution of the defects can be harnessed to use the scars as scaffolding that can template the assembly of functional nanomolecules or form the connectors for a space-filling lattice of functionalized spheres.    Instead, the colloidosome building blocks can be easily synthesized, controlled and assembled so that greater attention can be lavished upon the active chemical or biochemical sites.  Morever, the colloidosomes themselves can be filled and functionalized with a vast array of therapeutics, biomaterials and nanostructures, enabling the assembly of truly hierarchical, self-processed materials.

\listrefs

\end